\documentclass[aps,pra,floatfix,showpacs,twocolumn,superscriptaddress]{revtex4}

\usepackage{graphicx}
\usepackage{latexsym}

\newcommand{\la}{\langle}
\newcommand{\ra}{\rangle}
\newcommand{\tr}{\mbox{tr}}

\newcommand{\qed}{$\hfill \Box$}
\newcommand{\poly}{\mbox{poly}}
\newcommand{\vsprime}{\vec{s}\,'}

\newcommand{\swap}{\mbox{{\sc swap}}}

\newcommand{\ket}[1]{\mbox{$|#1\rangle$}}

\newtheorem{theorem}{Theorem}
\newtheorem{proposition}{Proposition}
\newtheorem{lemma}{Lemma}

\begin{document}

\title{How robust is a quantum gate in the presence of noise?}

\author{Aram W. Harrow}
\email{aram@mit.edu}
\homepage[\\ URL:]{http://web.mit.edu/aram/}
\affiliation{MIT Physics, 77 Massachusetts Ave., Cambridge MA 02139, USA}
\affiliation{School of Physical Sciences, University of Queensland,
Queensland 4072, Australia}
\author{Michael A. Nielsen}
\email{nielsen@physics.uq.edu.au}
\homepage[\\ URL:]{http://www.qinfo.org/people/nielsen/}
\affiliation{School of Physical Sciences, University of Queensland,
Queensland 4072, Australia}

\date{\today}

\begin{abstract}
  We define several quantitative measures of the \emph{robustness} of
  a quantum gate against noise.  Exact analytic expressions for the
  robustness against depolarizing noise are obtained for all unitary
  quantum gates, and it is found that the controlled-{\sc not} is the
  most robust two-qubit quantum gate, in the sense that it is the
  quantum gate which can tolerate the \emph{most} depolarizing noise
  and still generate entanglement.  Our results enable us to place
  several analytic upper bounds on the value of the threshold for
  quantum computation, with the best bound in the most pessimistic
  error model being $p_{\rm th} \leq 0.5$.
\end{abstract}


\maketitle

\section{Introduction}

%
%
An ideal quantum computer~\cite{Nielsen00a} is usually described as a
sequence of unitary quantum gates applied to the qubits making up the
computer.  A typical universal set of quantum gates is the
controlled-{\sc not} gate, and single-qubit unitary
operations~\cite{Barenco95a}.  A crucial element in a universal gate
set is that it be capable of generating \emph{entanglement} between
the qubits making up the computer.

%
%
In the real world quantum gates suffer from noise~\cite{Landauer95a},
which can inhibit the creation of entanglement.  This problem led to
the development of fault-tolerant methods for quantum computation (see
the discussion and references in~\cite{Nielsen00a}) based on quantum
error-correcting codes~\cite{Shor95a,Steane96c}.  One of the
outstanding achievements of work on fault-tolerance is the
\emph{threshold theorem} for quantum
computation~\cite{Aharonov99a,Gottesman97a,Kitaev97a,Knill98a,Preskill98b}.
The threshold theorem states that, under reasonable physical
assumptions about noise in the computer, it is possible to correct for
the effects of that noise, provided the strength of the noise is below
some constant threshold, $p_{\rm th}$.  (Roughly speaking, $p_{\rm
  th}$ can be thought of as the maximal probability of error during a
single quantum gate that can be corrected using the methods of
fault-tolerance.)  The exact value of the threshold depends on what
assumptions are made about the noise in the quantum computer, and
estimates of the value of the threshold therefore vary quite a bit.
Typical current estimates place it in the range $10^{-4}$ to
$10^{-6}$.

%
%
Motivated by the practical problem of noise, and the theory of
fault-tolerant quantum computation, in this paper we consider the
problem of quantifying how robust a quantum gate is to the effects of
noise.  More precisely, for a given gate $U$ we attempt to quantify
how much noise the gate can tolerate while preserving the ability to
generate entanglement.  Since, in a sense we make precise below,
entanglement generation is necessary for quantum computation to be
possible, even if the methods of fault-tolerant computation are used,
this program allows us to determine \emph{upper bounds} on the value
of the threshold.

%
%
Our work is different from most other work on estimating thresholds,
which usually aims to determine \emph{lower bounds}.  The interest in
lower bounds stems from their more immediate practical interest: if we
know that $p_{\rm th} > 10^{-6}$, for example, then that gives
experimentalists a target to shoot for in pursuit of a working quantum
computer.  Nonetheless, as emphasized in~\cite{Aharonov96b}, from a
fundamental point of view it would be extremely interesting to have
exact values for the threshold, and this requires techniques for
obtaining upper bounds.

%
%
Our work is based upon the results of Vidal and
Tarrach~\cite{Vidal99b}, who investigated the \emph{robustness} of
entangled quantum states, that is, how much noise can be added to a
quantum state before it becomes unentangled, i.e., separable.  Our
work also naturally extends and complements the work of Aharonov and
Ben-Or~\cite{Aharonov96b}, who, to our knowledge, have done the only
prior work obtaining upper bounds on the value of the threshold.

%
%
Another interesting context in which our measures of gate robustness
may be placed is the program of defining ``dynamic strength measures''
for quantum dynamical operations~\cite{Nielsen02a}.  Dynamic strength
measures quantify the intrinsic power or strength of a quantum
dynamical operation as a physical resource, much as an entanglement
measure quantifies the entanglement in a quantum state.
\cite{Nielsen02a} developed a framework for the analysis of dynamic
strength measures, and we will see that gate robustness can be
regarded as a measure of dynamic strength, and analyzed within this
framework.

%
%
The structure of the paper is as follows.  Sec.~\ref{sec:op-schmidt}
reviews background material on the Schmidt decomposition for
operators.  This decomposition is central to our later work on the
robustness of quantum gates.  Sec.~\ref{sec:separable} reviews the
notion of \emph{separable} quantum gates, which may be defined as the
class of gates that cannot generate entanglement in a quantum
computer.  Furthermore, this section proves that a quantum circuit
containing only separable gates can be efficiently simulated on a
classical computer.  Sec.~\ref{sec:robustness} reviews Vidal and
Tarrach's work on the robustness of quantum states.  This section also
introduces a novel measure of the robustness of quantum states useful
in our later work on gate robustness, and proves some elementary
properties of the new measure.  Sec.~\ref{sec:results} gives our
definitions and results on the robustness of quantum gates, and
relates the results to the theory of fault-tolerant quantum
computation.  Sec.~\ref{sec:conclusion} concludes.

\section{The operator-Schmidt decomposition}
\label{sec:op-schmidt}

%
%
The \emph{operator-Schmidt} decomposition is an operator analogue of
the well-known Schmidt decomposition for pure quantum
states~\cite{Nielsen00a}.  The present treatment of the
operator-Schmidt decomposition is based on the discussion
in~\cite{Nielsen02a,Nielsen98d}, with the addition of a result on the
continuity of the Schmidt coefficients of a unitary operator.

%
%
We begin by introducing the Hilbert-Schmidt inner product on $d\times
d$ operators, $(Q,P)\equiv\mbox{tr}(Q^\dag P)$, for any operators $Q$
and $P$.  We define an orthonormal operator basis to be a set
$\{Q_j\}$ which satisfies the condition $(Q_j,Q_k)=\tr(Q_j^\dag
Q_k)=\delta_{jk}$.  For example, an orthonormal basis for the space of
single-qubit operators is the set
$\{I/\sqrt2,X/\sqrt2,Y/\sqrt2,Z/\sqrt2\}$, where $X$, $Y$, and $Z$ are
the Pauli sigma operators, and $I$ is the identity.

%
%
The operator-Schmidt decomposition states that any operator $Q$ acting
on systems $A$ and $B$ may be written~\cite{Nielsen98d}:
\begin{eqnarray} \label{eq:operator_schmidt}
Q=\sum_l q_l A_l\otimes B_l,
\end{eqnarray}
where $q_l\geq 0$, and $A_l$ and $B_l$ are orthonormal operator bases
for $A$ and $B$, respectively.  To prove the operator-Schmidt
decomposition, expand $Q$ in the form $Q=\sum_{jk}M_{jk}C_j\otimes
D_k$, where $C_j$ and $D_k$ are fixed orthonormal operator bases for
$A$ and $B$, respectively, and $M_{jk}$ are complex coefficients.  The
singular value decomposition states that the matrix $M$ with $(j,k)$th
entry $M_{jk}$ may be written $M=UqV$, where $U$ and $V$ are unitary
matrices and $q$ is a diagonal matrix with non-negative entries.  We
thus obtain
\begin{eqnarray}
Q=\sum_{jkl}U_{jl}q_l V_{lk}C_j\otimes D_k,
\end{eqnarray}
where $q_l$ is the $l$th diagonal entry of $q$.  Defining orthonormal
operator bases $A_l\equiv \sum_j U_{jl}C_j$ and $B_l\equiv \sum_k
V_{lk}D_k$, we obtain the operator-Schmidt decomposition,
Eq.~(\ref{eq:operator_schmidt}).

%
%
To better understand the coefficients $q_l$ in the operator-Schmidt
decomposition, imagine that associated with each system, $A$ and $B$,
there are \emph{reference systems}, $R_A$ and $R_B$, with the same
state space dimensionalities, $d_A$ and $d_B$, as $A$ and $B$.  Let
\begin{eqnarray}
|\alpha\ra & \equiv & \sum_j |j_{R_A} j_A\ra / \sqrt{d_A}, \,\, 
\rm{  and } \\
|\beta\ra & \equiv & \sum_j |j_B j_{R_B}\ra / \sqrt{d_B}
\end{eqnarray}
denote normalized, maximally entangled states of $R_A A$ and $B R_B$,
respectively.  Now let ${\cal E}$ be a general quantum
operation~\footnote{Quantum operations are sometimes known as
  \emph{completely positive maps}.  We use the more physically
  oriented terminology, since it is physical applications we have in
  mind.  Note that we use ``quantum gate'' and ``quantum operation''
  interchangeably, depending on whether the context is quantum
  computation or more general.  A review of the theory of quantum
  operations may be found in~\cite{Nielsen00a}.}; we will shortly
specialize to the case when ${\cal E}$ corresponds to the action of
$U$.  We define $\rho({\cal E})$ to be the density operator resulting
when ${\cal E}$ acts on $|\alpha\ra|\beta\ra$.  Writing this out
explicitly, with subscripts to make it clear which operations are
acting on which systems:
\begin{eqnarray} \label{eq:rho-e}
 \rho({\cal E}) & \equiv & \left( {\cal I}_{R_A} \otimes {\cal E}_{AB} 
    \otimes {\cal I}_{R_B}\right) \circ \left(|\alpha\ra \la \alpha| \otimes
     |\beta \ra \la \beta|\right),
\end{eqnarray}
where ${\cal I}_S$ denotes the identity quantum operation on a system
$S$.  In the special case when ${\cal E}$ represents a unitary
operation, $U$, on $AB$, we define $\psi(U)$ to be the quantum state
obtained when $U$ acts on $|\alpha\rangle |\beta\rangle$, and let
$\rho(U)$ be the corresponding density operator.  Note that we will
interchange notations like $\psi(U)$ and $|\psi(U)\rangle$, depending
on which is more convenient in a particular context.

%
%
The Schmidt coefficients of $\psi(U)$ are closely connected to the
operator-Schmidt coefficients of $U$, which we denote $u_j$.
Letting $U = \sum_j u_j A_j \otimes B_j$ be an operator-Schmidt decomposition,
we see that
\begin{eqnarray}
  \psi(U) & = & (I_{R_A} \otimes U \otimes I_{R_B}) 
  |\alpha\rangle |\beta\rangle
  \\
  & = & \sum_j u_j (I_{R_A} \otimes A_j) |\alpha\rangle 
  (B_j \otimes I_{R_B})|\beta\rangle.
\end{eqnarray}
Direct calculation shows that $\sqrt{d_A}(I_{R_A} \otimes A_j)
|\alpha\rangle$ and $\sqrt{d_B}(B_j \otimes I_{R_B})|\beta\rangle$ form
orthonormal bases for $R_AA$ and $BR_B$, respectively.  Thus, we
obtain the useful result that the quantum state $\psi(U)$ has Schmidt
coefficients $u_j/\sqrt{d_A d_B}$ equal, up to the factor $1/\sqrt{d_A
  d_B}$, to the Schmidt coefficients of $U$.

%
%
The following proposition shows that the Schmidt coefficients of $U$
are continuous functions of $U$.  In the statement of the proposition,
$\| M \| = \max_{\| \psi \| = 1} \| M|\psi\rangle \|$ denotes the
usual operator norm.

\begin{proposition}{\label{prop:Schmidt-continuity}}
  Let $U$ and $V$ be operators on $AB$, with respective Schmidt
  coefficients $u_j$ and $v_j$, ordered into decreasing order, $u_1
  \geq u_2 \geq \ldots$, and $v_1 \geq v_2 \geq \ldots$.  Then
  \begin{eqnarray} \label{eq:Schmidt-continuity}
    2 \left(1-\frac{\sum_j u_j v_j}{d_Ad_B}\right) \leq \| U-V \|^2
  \end{eqnarray}
\end{proposition}

To understand why Eq.~(\ref{eq:Schmidt-continuity}) can be interpreted
as a statement about continuity requires a little thought.  Note that
$\tr(U^\dagger U) = \tr(V^\dagger V) = d_A d_B$, and thus $\sum_j
u_j^2 = \sum_j v_j^2 = d_A d_B$.  It follows that we can think of
$u_j^2/d_A d_B$ and $v_j^2/d_A d_B$ as probability distributions.
With this interpretation, the quantity $\sum_j u_jv_j / d_A d_B$ is
just the fidelity of these two probability distributions, and it
follows from Eq.~(\ref{eq:Schmidt-continuity}) that if $U \approx V$
then $u_j \approx v_j$ for all $j$.

\textbf{Proof:} The key is to observe that the norm $\| \cdot \|$ is
\emph{stable} when extended trivially to an ancilla system, i.e., $\|
M \| = \|M \otimes I\|$.  Using this observation we have
\begin{eqnarray}
  \| U-V\| & = & \| I_{R_A} \otimes (U-V) \otimes I_{R_B} \| \\
  & \geq & \| \left( I_{R_A} \otimes (U-V) \otimes I_{R_B} 
    \right) |\alpha \rangle |\beta \rangle \|  \\
    & = & \| \psi(U) - \psi(V) \|.
\end{eqnarray}
Squaring both sides of the inequality, and interchanging the roles of
the two sides, we obtain:
\begin{eqnarray}
  \| \psi(U) \|^2 + \| \psi(V) \|^2 -2 \, \mbox{Re}\left( 
  \langle \psi(U)|\psi(V)\rangle \right)
  \leq \| U-V\|^2. \nonumber \\
\end{eqnarray}
Since $\| \psi(U)\|^2 = \| \psi(V) \|^2 = 1$, this implies
\begin{eqnarray} \label{eq:cont-inter}
  2\left( 1-|\langle \psi(U)|\psi(V) \rangle |\right)
  \leq \| U-V\|^2.
\end{eqnarray}
Since $\psi(U)$ and $\psi(V)$ have Schmidt coefficients
$u_j/\sqrt{d_Ad_B}$ and $v_j/\sqrt{d_A d_B}$, respectively, it follows
from the results of~\cite{Barnum99a,Vidal00a} that $|\langle
\psi(U)|\psi(V)\rangle| \leq \sum_j u_j v_j / d_A d_B$.  Combining
this inequality with Eq.~(\ref{eq:cont-inter}) gives the desired
result.
 \qed

\section{Separable and separability-preserving quantum gates}
\label{sec:separable}

We now formally introduce the notion of separable quantum gates, and
study their basic properties, in Sec.~\ref{subsec:sep-basic}.
Sec.~\ref{subsec:sep-simulate} states and proves a theorem showing
that quantum circuits built entirely out of separable quantum gates
can be efficiently simulated on a classical computer.  Finally,
Sec.~\ref{subsec:sep-preserving} notes that the classical simulation
theorem of the previous subsection can be extended to a somewhat
larger class of gates, the ``separability-preserving'' gates, and
considers some of the implications of this fact.

\subsection{Definition and basic properties}
\label{subsec:sep-basic}

%
%
Suppose ${\cal E}$ is a quantum operation acting on a composite
quantum system with two components labeled $A$ and $B$.  ${\cal E}$ is
said to be \emph{separable} if it can be given an operator-sum
representation of the form
\begin{eqnarray}
  {\cal E}(\rho) = \sum_j (A_j \otimes B_j) \rho 
(A_j^\dagger \otimes B_j^\dagger)
\end{eqnarray}
Separable quantum operations were independently introduced
in~\cite{Vedral97a,Barnum98a}, where it was speculated that
trace-preserving separable quantum operations might correspond to the
class of quantum operations that can be implemented on a bipartite
system using local operations and classical communication.  This
speculation was false~\cite{Bennett99d}.  However, a related
conjecture is true, namely, that trace-preserving separable quantum
operations correspond to the class of trace-preserving quantum
operations which cannot be used to generate quantum entanglement.
This follows from an elegant characterization theorem of Cirac
\emph{et al}~\cite{Cirac01a} linking separability of a quantum
operation ${\cal E}$ to separability of the quantum state $\rho({\cal
  E})$ introduced in Eq.~(\ref{eq:rho-e}).

\begin{theorem}[Operation-separability theorem 
\cite{Cirac01a}]{\label{thm:op-sep}} A trace-preserving
  quantum operation ${\cal E}$ is separable if and only if $\rho({\cal
    E})$ is a separable quantum state, that is, $\rho({\cal E})$ can
  be written in the form
\begin{eqnarray}
  \rho({\cal E}) = \sum_j p_j \rho^{R_AA}_j \otimes \rho^{BR_B}_j,
\end{eqnarray}
where the $p_j$ are probabilities, $\rho^{R_AA}_j$ are quantum states
of system $R_AA$, and $\rho^{BR_B}_j$ are quantum states of system
$BR_B$.
\end{theorem}

%
%
When we say in the statement of the theorem that $\rho({\cal E})$ is
separable there is initially some ambiguity, due to the multiple ways
the system $R_AABR_B$ can be decomposed into subsystems.  To avoid
this ambiguity, it is convenient to introduce notational conventions
as follows.  Let $\sigma$ be a state of a composite system $CD$.  We
say $\sigma$ is \emph{separable with respect to the $C:D$ cut} if
$\sigma$ can be written $\sigma = \sum_j p_j \rho^C_j \otimes
\rho^D_j$ for probabilities $p_j$, and quantum states $\rho^C_j,
\rho^D_j$ of systems $C$ and $D$, respectively.  The advantage of this
notation comes when more systems are introduced.  For example, when
$\sigma$ is a state of a tripartite system, $CDE$, it is immediately
clear what we mean by separability with respect to the $C:DE$ cut, or
with respect to the $C:D:E$ cut, or other possible cuts.  Thus, in the
operation-separability theorem, the assertion is that ${\cal E}$ is
separable if and only if $\rho({\cal E})$ is separable with respect to
the $R_AA:BR_B$ cut.

%
%
We have stated the operation-separability theorem for the case of
trace-preserving quantum operations, but a similar result also holds
for non trace-preserving quantum operations ${\cal E}$.  The only
change is that the $p_j$ are no longer probabilities, but instead can
be any set of non-negative real numbers.  We have also restricted our
attention to bipartite quantum operations, that is, ${\cal E}$ which
act on quantum systems with just two components, $A$ and $B$.  It is
not difficult to show that an analogous statement holds for $k$-party
quantum operations ${\cal E}$.  We do this by endowing each party with
an associated reference system with which it is initially maximally
entangled, and defining $\rho({\cal E})$ to be the result of allowing
${\cal E}$ to act on this initial state.  ${\cal E}$ is then separable
if and only if $\rho({\cal E})$ is separable.

%
%
An interesting corollary of the operation-separability theorem is that
a quantum operation is separable if and only if it is incapable of
producing entangled states.  Furthermore, by connecting gate
separability to state separability, the operation-separability theorem
allows us to apply results from the theory of state separability to
prove that certain gates are separable, and thus incapable of
producing entanglement.

%
%
The operation-separability theorem tells us that a trace-preserving
quantum operation ${\cal E}$ is separable precisely when $\rho({\cal
  E})$ is separable.  However, it does not follow that all separable
states of $R_AA:BR_B$ can be written as $\rho({\cal E})$ for some
trace-preserving quantum operation.  To understand why this is the
case, observe that when ${\cal E}$ is trace-preserving,
$\tr_{AB}(\rho({\cal E}))$ must be the completely mixed state of $R_A
R_B$.  In general, however, it is not difficult to find separable
states $\sigma$ of $R_AA:BR_B$ such that $\tr_{AB}(\sigma)$ is not
completely mixed.

%
%
An elegant result of M.,~P., and~R.~Horodecki~\cite{Horodecki99c} can
be used to characterize precisely which separable states can be
written in the form $\rho({\cal E})$ for trace-preserving, separable
${\cal E}$.  Their result, which we have restated in the context of
multipartite systems, is as follows:

\begin{theorem}
  The set of density matrices, $\sigma$, of $R_AABR_B$ such that
  $\sigma = \rho({\cal E})$ for some trace-preserving quantum
  operation ${\cal E}$ is precisely the set such that
  $\tr_{AB}(\sigma)$ is the completely mixed state of $R_AR_B$.
\end{theorem}

%
%
Combining this theorem with the operation-separability theorem we
obtain the following result:
\begin{theorem}{\label{thm:sep-char}}
  The set of density matrices, $\sigma$, of $R_AABR_B$ such that
  $\sigma = \rho({\cal E})$ for some trace-preserving and separable
  quantum operation ${\cal E}$ is precisely the set such that (a)
  $\sigma$ is separable with respect to the $R_AA:BR_B$ cut; and (b)
  $\tr_{AB}(\sigma)$ is the completely mixed state of $R_AR_B$.
\end{theorem}

\subsection{Separable gates and quantum computation}
\label{subsec:sep-simulate}

%
%
Having discussed the basic properties of separable quantum operations,
we now turn to their utility for quantum computation.  Imagine a
quantum circuit is built entirely out of separable quantum gates and
single-qubit gates.  It is intuitively plausible that such a quantum
circuit can be efficiently simulated on a classical computer, and we
now prove this result.  The major technical difficulty is issues
involving the accuracy required in the simulation, and the associated
computational overhead.

%
%
Our model of quantum computation is as follows.  Let ${\cal G}$ be a
fixed set of one- and two-qubit quantum gates.  By ``quantum gate'' we
mean a trace-preserving quantum operation.  We assume that all the
two-qubit gates in ${\cal G}$ are separable.  We let $\{ C_n \}$ be a
uniform family of quantum circuits~\cite{Nielsen00a,Yao93a} containing
$p(n)$ gates, and acting on $q(n)$ qubits, where $p(n)$ and $q(n)$ are
polynomials in some parameter $n$.  The initial state of the computer
is assumed to be a computational basis state, $|x\rangle$.  The
computation is concluded by performing a measurement in the
computational basis, yielding a probability distribution $p_x(y)$ over
possible measurement outcomes $y$.  The measurement may be either on
all the qubits, or on some prespecified subset.  For instance, if one
is solving a decision problem, it is only necessary to measure the
first qubit of the computer, to get a single zero or one as output.

%
%
What does it mean to simulate this computation efficiently on a
classical computer?  Suppose we have a classical computer that, on
input of $x$, produces an output $y$ with probability distribution
$\tilde p_x(y)$.  A good measure of how well this simulates the
quantum computation is provided by the \emph{$L_1$ distance}.  For
probability distributions $r(y)$ and $s(y)$ the $L_1$ distance is
defined by $D(r(y),s(y)) \equiv \sum_{y} |r(y)-s(y)|/2$.  Thus, we
require that the $L_1$ distance $D(p_x(y),\tilde p_x(y)) = \sum_y
|p_x(y)-\tilde p_x(y)|/2$ satisfies
\begin{eqnarray}
D(p_x(y),\tilde p_x(y)) \leq \epsilon
\end{eqnarray}
for some parameter $\epsilon > 0$.  We will show that the
computational resources required to achieve this accuracy on a
classical computer scale as $O(\poly(p(n)/\epsilon))$, where
$\poly(\cdot)$ is some polynomial of fixed degree not depending on the
circuit family $\{ C_n \}$.  Thus, high accuracies in the simulation
can be achieved with modest computational cost.

%
%
As an example of the practical implications of this result, suppose
$\{ C_n \}$ is a uniform family of quantum circuits solving a decision
problem, outputting the correct answer to an instance, $x$, of the
decision problem with probability at least $3/4$.  Our result implies
that there is a classical simulation using $O(\poly(p(n)))$ gates, and
outputting the correct solution to the decision problem with
probability $2/3$. (The probability of obtaining the correct answer
may easily be boosted up beyond $3/4$ by a constant number of
repetitions.)

%
%
To analyze the method described below for classical simulation, we
need the notion of the \emph{trace distance}, a quantum generalization
of the $L_1$ distance.  The trace distance, $D(\rho,\sigma)$, between
density matrices $\rho$ and $\sigma$ is defined by~\cite{Nielsen00a}
$D(\rho,\sigma) \equiv \tr|\rho-\sigma|/2$.  Note that we use the same
notation $D(\cdot,\cdot)$ for the trace distance and the $L_1$
distance, with the meaning to be determined from context.  The
properties of the trace distance are discussed in detail
in~\cite{Nielsen00a}, and we need only a few properties here:
\begin{itemize}
\item The trace distance satisfies the \emph{triangle inequality},
$D(\rho,\tau) \leq D(\rho,\sigma)+D(\sigma,\tau)$.

\item The trace distance is \emph{doubly convex}, meaning that if
  $p_j$ are probabilities, and $\rho_j$ and $\sigma_j$ are
  corresponding density matrices, then
  \begin{eqnarray}
    D\left( \sum_j p_j \rho_j,\sum_j p_j \sigma_j\right) \leq 
    \sum_j p_j D(\rho_j,\sigma_j).
  \end{eqnarray}

\item The trace distance is \emph{contractive}.  That is, if ${\cal
    E}$ is a trace-preserving quantum operation, then
  $D({\cal E}(\rho),{\cal E}(\sigma)) \leq D(\rho,\sigma)$.

\item The trace distance has the \emph{stability property}
$D(\rho_1 \otimes \sigma, \rho_2 \otimes \sigma) = D(\rho_1,\rho_2)$.

\item Suppose $E_y$ are POVM elements describing the statistics from
  an arbitrary quantum measurement.  Let $r(y) \equiv \tr(\rho E_y)$
  and $s(y) \equiv \tr(\sigma E_y)$ be the corresponding probability
  distributions for $\rho$ and $\sigma$.  Then the $L_1$ distance and
  the trace distance are related by the inequality
  \begin{eqnarray} \label{eq:t-d-inequality}
    D(r(y),s(y)) \leq D(\rho,\sigma).
  \end{eqnarray}

\end{itemize}

We now describe how the classical simulation is performed, followed by
an analysis to determine the accuracy of the simulation. 

\textbf{Variables used in the classical simulation:} For each
$j=1,\ldots,q(n)$ we let $\vec s_j$ be a three-dimensional real
vector.  Each vector $\vec s_j$ is \emph{valid}, meaning that it has
the following three properties: (a) Each component of $\vec s_j$ is in
the range $[-1,1]$; (b) Each component is specified to $l$ bits of
precision, where $l$ is a number that will be fixed by the later
analysis, in order to ensure the overall accuracy is at least
$\epsilon$; and (c) $\| \vec s_j \| \leq 1$.

%
%
We use the notation $\vec s \equiv (\vec s_1,\ldots,\vec s_{q(n)})$ to
denote the $3q(n)$-dimensional real vector containing all the $\vec
s_j$s as subvectors.  We say that $\vec s$ is \emph{valid} if each
$\vec s_j$ is valid.  It will also be convenient to introduce the
notation
\begin{eqnarray}
  \rho(\vec s) \equiv \frac{I+\vec s_1 \cdot \vec \sigma}{2} \otimes \ldots
  \otimes \frac{I+\vec s_{q(n)} \cdot \sigma}{2}.
\end{eqnarray}
Note that $\rho(\vec s)$ is a legitimate density operator of $q(n)$
qubits, whenever $\vec s$ is valid.  The idea of the classical
simulation is that the variables $\vec s$ will be used to represent
the state $\rho(\vec s)$.  Note that $\rho(\vec s)$ is \emph{not} a
variable used in the classical simulation; it is simply a mathematical
notation convenient in the analysis of the simulation.

\textbf{Initial state of the classical variables:} Suppose the initial
state of the quantum computer is $|x\rangle$, where $x$ has binary
expansion $x_1\ldots x_{q(n)}$.  If $x_j = 0$ we set $\vec s_j =
(0,0,1)$ initially, while if $x_j = 1$ we set $\vec s_j = (0,0,-1)$
initially.

\textbf{Simulating a single-qubit gate:} A single-qubit gate can be
regarded as a two-qubit separable gate in which one of the qubits is
acted on trivially.  Thus, we need only consider the case of two-qubit
separable gates.

\textbf{Simulating a two-qubit separable gate:} Suppose ${\cal E}$ is
a two-qubit separable gate, and it acts on qubits $A$ and $B$.  We
simulate this gate by using $\vec s$ as input to the following
stochastic gate simulation procedure, which produces a valid
$3q(n)$-dimensional vector, $\vsprime$, as output.  We then set $\vec
s = \vsprime$, and repeat over, going through each gate, ${\cal
  E}_1,\ldots, {\cal E}_{p(n)}$, in the computation, until a final
output value of $\vec s$ is produced, at which point we proceed to the
simulation of the final measurement, as described below.

\textbf{Gate simulation procedure:} 

\textit{Input to the procedure:} A valid vector, $\vec s$.

\textit{Body of the procedure:} Find valid three-vectors $\vec
s^{j}_A$ and $\vec s^{j}_B$, a probability distribution, $p_j$,
containing at most $16$ elements, and with each $p_j$ specified to $l$
bits of precision, such that
\begin{eqnarray}
  & & D\Bigg( {\cal E}
    \left(\frac{I+\vec s_A \cdot \vec \sigma}{2} \otimes
      \frac{I+\vec s_B \cdot \vec \sigma}{2} \right),  \nonumber \\
  & & \hphantom{D\Bigg( \,\, }
    \sum_j p_j 
    \frac{I+\vec s^{j}_A \cdot \vec \sigma}{2} \otimes
    \frac{I+\vec s^{j}_B \cdot \vec \sigma}{2} \Bigg)
  \leq c 2^{-l}, \label{eq:gate-sim}
\end{eqnarray}
for some constant $c$ that does not depend on ${\cal E}, A$ or $B$.
To see that this is possible, we make use of the fact that
\begin{eqnarray}
{\cal E} \left(\frac{I+\vec s_A \cdot \vec \sigma}{2} \otimes
      \frac{I+\vec s_B \cdot \vec \sigma}{2} \right)
\label{eq:sep-input}
\end{eqnarray}
is a separable, two-qubit state, and therefore, by Carath\'{e}odory's
theorem~\cite{Rockafeller70a}, can be written in the form
\begin{eqnarray}
   \sum_j q_j 
    \frac{I+\vec t^{j}_A \cdot \vec \sigma}{2} \otimes
    \frac{I+\vec t^{j}_B \cdot \vec \sigma}{2},
\end{eqnarray}
where the $q_j$ are probabilities, $\vec t^j_A, \vec t^j_B$ are
real-three vectors satisfying $\| \vec t^j_A \|, \| \vec t^j_B\| \leq
1$, and there are at most $16$ terms in the sum.  Choosing the $p_j$
to be probabilities which are $l$-bit approximations to the $q_j$, and
the $\vec s^j_A, \vec s^j_B$ to be valid vectors which approximate
$\vec t^j_A, \vec t^j_B$ also to $l$ bits, we obtain the result.

%
%
Note that while Carath\'{e}odory's theorem ensures that such
probabilities and vectors exist, finding them may be a non-trivial
task.  The obvious technique, a brute force search over probability
distributions and valid vectors, requires $\poly(2^l)$ operations,
where $\poly(\cdot)$ is some fixed polynomial function.  Although we
believe it likely that better techniques --- perhaps even polynomial
in $l$ --- are possible, for the purposes of the present simulation
$\poly(2^l)$ turns out to be sufficient.

\textit{Output of the procedure:} For $k \neq A, B$ we define $\vec
s^{j}_k \equiv \vec s_k$.  Set $\vec s^j = (\vec s^j_1,\ldots,\vec
s^j_{q(n)})$.  Note that $\vec s^j$ is valid, by construction.  With
probability $p_j$, output $\vsprime = \vec s^j$.

\textbf{Simulating the final measurement in the computational basis:}
Let $S$ be the subset of qubits that is measured at the output of the
quantum computation.  For each $k \in S$, let $s^3_k$ be the third
component of $\vec s_k$.  The measurement result for that qubit is $0$
with probability $(1+s^3_k)/2$, and $1$ with probability
$(1-s^3_k)/2$.  Note that, by definition, $\tilde p_x(y)$ is the
distribution over possible outcomes, $y$, produced by following this
procedure.

\textbf{Analysis:} The key to the analysis of the classical simulation
is a simple equivalence between the classical simulation and certain
measurements on quantum states.  Suppose we define $\tilde p^m(\vec
s)$ to be the probability distribution on valid vectors after $m$
steps of the simulation procedure, that is, after ${\cal
  E}_1,\ldots,{\cal E}_m$ have been simulated.  For $m = 0,\ldots,
p(n)$ define
\begin{eqnarray}
  \tilde \sigma^m \equiv \sum_{\vec s} \tilde p^m(\vec s) \rho(\vec s).
\end{eqnarray}
It is not difficult to see that the distribution obtained by measuring
$\tilde \sigma^{p(n)}$ in the computational basis of the subset $S$ is
exactly the same as the output distribution $\tilde p_x(y)$ produced
by the classical simulation.

%
%
For $m = 0,\ldots,p(n)$ define $\sigma^m$ to be the state of the
actual quantum computer after $m$ gates have been applied.  Thus
$\sigma^0 = |x\rangle \langle x|$, $\sigma^1 = {\cal E}_1(\sigma^0)$,
and so on.  The idea of the proof that the classical simulation works
well is to bound the distance between $\sigma^m$ and $\tilde
\sigma^m$.  We do this using the following lemma.

\begin{lemma}{\label{lemma:simulation}}
  Suppose a valid vector $\vec s$ is used as input to the gate
  simulation procedure with probability $p(\vec s)$, and let
  $p(\vsprime)$ be the corresponding output distribution on valid
  vectors.  Define
  \begin{eqnarray}
    \sigma & \equiv & \sum_{\vec s} p(\vec s) \rho(\vec s) \\
    \sigma' & \equiv & \sum_{\vsprime} p(\vsprime) \rho(\vsprime)
  \end{eqnarray}
  If the gate simulation procedure simulates the gate ${\cal E}$, then
  we have
\begin{eqnarray}
  D({\cal E}(\sigma),\sigma') \leq c2^{-l},
\end{eqnarray}
where $c$ is the constant introduced earlier in the discussion of the gate
simulation procedure.
\end{lemma}

\textbf{Proof:} Let $p(\vsprime|\vec s)$ be the probability that
$\vsprime$ is output by the gate simulation procedure, given that
$\vec s$ is input.  Then we have $p(\vsprime) = \sum_{\vec s}
p(\vsprime| \vec s) p(\vec s)$, so
\begin{eqnarray}
  \sigma' & = & \sum_{\vec s} p(\vec s) \sum_{\vsprime} p(\vsprime|\vec s)
  \rho(\vsprime).
\end{eqnarray}
Applying the double convexity of the trace distance gives
\begin{eqnarray} \label{eq:lemma}
  D({\cal E}(\sigma),\sigma') \leq
  \sum_{\vec s} p(\vec s) D\left({\cal E}(\rho(\vec s)),\sum_{\vsprime}
  p(\vsprime|\vec s) \rho(\vsprime)\right). \nonumber \\
\end{eqnarray}
By inspection of the construction used in the gate simulation
procedure, notably Eq.~(\ref{eq:gate-sim}), and the stability property
for trace distance, we have
\begin{eqnarray}
  D\left({\cal E}(\rho(\vec s)),\sum_{\vsprime}
  p(\vsprime|\vec s) \rho(\vsprime)\right) \leq c2^{-l}.
\end{eqnarray}
Combining this observation with Eq.~(\ref{eq:lemma}) gives
\begin{eqnarray}
  D({\cal E}(\sigma),\sigma') \leq c2^{-l},
\end{eqnarray}
which was the desired result. \qed

\begin{proposition}
For $m=0,\ldots,p(n)$, $D(\sigma^m,\tilde \sigma^m) \leq cm2^{-l}$.
\end{proposition}

\textbf{Proof:} We induct on $m$.  For $m = 0$ the result follows from
the fact that $\sigma^0 = \tilde \sigma^0$.  Assuming the result is
true for $m$, we now prove it for $m+1$.  By the triangle inequality
\begin{eqnarray}
  D(\sigma^{m+1},\tilde \sigma^{m+1}) & \leq & D(\sigma^{m+1},
  {\cal E}_{m+1}(\tilde \sigma^m)) \nonumber \\
  & & +D({\cal E}_{m+1}(\tilde \sigma^m),\tilde \sigma^{m+1}). 
\end{eqnarray}
By definition $\sigma^{m+1} = {\cal E}_{m+1}(\sigma^m)$, so this
equation may be rewritten
\begin{eqnarray}
  D(\sigma^{m+1},\tilde \sigma^{m+1}) & \leq & D({\cal E}_{m+1}(\sigma^m),
  {\cal E}_{m+1}(\tilde \sigma^m)) \nonumber \\
  & & +D({\cal E}_{m+1}(\tilde \sigma^m),\tilde \sigma^{m+1}). 
\end{eqnarray}
Applying the contractivity of the trace distance to the first term,
and Lemma~\ref{lemma:simulation} to the second term, we obtain
\begin{eqnarray}
  D(\sigma^{m+1},\tilde \sigma^{m+1}) \leq D(\sigma^m,
  \tilde \sigma^m)
  +c2^{-l}.
\end{eqnarray}
Applying the inductive hypothesis to the first term gives
\begin{eqnarray}
  D(\sigma^{m+1},\tilde \sigma^{m+1}) \leq cm2^{-l}+c2^{-l} = c(m+1)2^{-l},
\end{eqnarray}
which completes the induction. \qed

%
%
We conclude from the proposition that $D(\sigma^{p(n)},\tilde
\sigma^{p(n)}) \leq cp(n) 2^{-l}$.  It follows from
Eq.~(\ref{eq:t-d-inequality}) that the simulated distribution
$\tilde p_x(y)$ and the actual distribution $p_x(y)$ are related by
the inequality
\begin{eqnarray}
  D(p_x(y),\tilde p_x(y)) \leq cp(n) 2^{-l}.
\end{eqnarray}
Choosing $l \equiv \lceil \log_2(cp(n)/\epsilon) \rceil$ we therefore have
\begin{eqnarray}
  D(p_x(y),\tilde p_x(y)) \leq \epsilon.
\end{eqnarray}
The total number of times the gate simulation procedure is performed
is $p(n)$, and the number of operations performed in one iteration of
the gate simulation procedure scales as $\poly(2^l)$, so the total
number of operations in the classical simulation is
$O(\poly(p(n)/\epsilon))$, where we abuse notation by letting
$\poly(\cdot)$ be a (new) polynomial function.  We have proved the
following theorem:

\begin{theorem}{\label{thm:simulation}}
  Let ${\cal G}$ be a fixed set of one- and two-qubit gates.  Suppose
  all two-qubit gates in ${\cal G}$ are separable.  Let $\{ C_n\}$ be
  a uniform family of quantum circuits of size $p(n)$, acting on
  $q(n)$ qubits, where both $p(n)$ and $q(n)$ are polynomials.  The
  initial state of the computer is assumed to be a computational basis
  state, $|x\rangle$.  The computation is concluded by performing a
  measurement in the computational basis on some prespecified subset,
  $S$, of the qubits, yielding a probability distribution $p_x(y)$
  over possible measurement outcomes $y$.  Then for any $\epsilon > 0$
  it is possible to sample from a distribution $\tilde p_x(y)$
  satisfying $D(p_x(y),\tilde p_x(y)) < \epsilon$ using a classical
  algorithm taking $O(\poly(p(n)/\epsilon))$ steps, where
  $\poly(\cdot)$ is some fixed polynomial.
\end{theorem}

%
%
Results related to Theorem~\ref{thm:simulation} have been obtained in
the past, but, so far as we have determined, no proof of this result
has previously been published.  In particular, Aharonov and
Ben-Or~\cite{Aharonov96b} studied the role of entanglement in quantum
computation, proving that many-party entanglement must be present in
order for a quantum computation to be difficult to simulate
classically.  This conclusion was subsequently clarified and extended
by Jozsa and Linden~\cite{Jozsa02a}.  However, the conclusions of
both~\cite{Aharonov96b} and~\cite{Jozsa02a} are not applicable in the
present context, since they apply in the context of pure state
entanglement of a quantum computer, rather than the mixed state case
considered in this paper.

%
%
The issue of mixed state quantum information processing was considered
by Braunstein \emph{et al}~\cite{Braunstein99a}, who raised, without
answering, the question of what role mixed-state entanglement can play
in quantum computation.  This line of thought has been carried further
by many authors, without completely answering the question.
See~\cite{Menicucci02a,Linden01a} for recent work and further
references.

\subsection{Separability-preserving gates}
\label{subsec:sep-preserving}

It is straightforward to extend the proof of
Theorem~\ref{thm:simulation} in a variety of ways, without changing
the conclusion that a classical simulation of the quantum circuit is
possible.  In particular, we can change the gates in ${\cal G}$ so
they can act on any \emph{bounded number} of \emph{qudit} systems,
rather than \emph{two-qubit} systems.  

Furthermore, the proof relies on properties of gates in ${\cal G}$
that are weaker than separability.  In particular, the gates in ${\cal
  G}$ need only be {\em separability-preserving}, that is, ${\cal
  E}(\rho)$ is separable for any separable state $\rho$.  We denote
the class of separability-preserving gates by {\bf SP}.  To see that
this is a weaker property, note that {\sc swap} is
separability-preserving since it maps product states to product
states, but {\sc swap} is not separable, since it can generate
entanglement with the aid of local ancilla systems.  More generally,
note that ${\cal E}_{AB}$ is separable with respect to $A:B$ if and
only if ${\cal E}_{AB}\otimes {\cal I}_{A'B'}$ is separability-preserving
with respect to $AA':BB'$.

Since the proof of Theorem~\ref{thm:simulation} only relied on the
state in Eq.~(\ref{eq:sep-input}) being separable, it still holds when
the available gates are all separability-preserving.  However, no
simple and easy-to-use characterization of the separability-preserving
gates is known, which is why we prefer, for most of the remainder of
this paper, to work with the separable gates.  We do make occasional
later use of separability-preserving gates, so it is convenient to
note here a few of their properties.  Note that all separable gates
are in {\bf SP}, and for gates operating on multiple qudits, any
permutation of the qudits (for example {\sc swap}) is in {\bf SP}.
Furthermore, {\bf SP} is convex and is closed under composition, so
\begin{eqnarray}
{\bf SP} \supseteq \mbox{Hull}\left\{ {\cal E} \circ {\cal P} :
{\cal E}\mbox{ separable and } {\cal P}\mbox{ a
permutation}\right\}. \nonumber \\
\end{eqnarray}
However, it is unclear whether this convex hull describes all of {\bf
SP}.  For example, the operation which measures a pair of a qubits in
the Bell basis and stores the answer in the computational basis (i.e.
$(\ket{00}+\ket{11})/\sqrt{2}$ becomes $\ket{00}$,
$(\ket{00}-\ket{11})/\sqrt{2}$ becomes $\ket{01}$, etc\ldots) is
certainly in {\bf SP} though is does not seem as though it can be
expressed as a convex combination of ${\cal E}_k\circ {\cal P}_k$ for
separable ${\cal E}_k$ and permutations ${\cal P}_k$~\footnote{We
thank Keiji Matsumoto for pointing this out to us.}.


\section{Robustness of quantum states}
\label{sec:robustness}

%
%
To understand how robust quantum gates are to noise, it is useful to
first review prior work on the robustness of entangled quantum states.
This section describes Vidal and Tarrach's~\cite{Vidal99b} definitions
and results on the robustness of quantum states, introduces a novel
measure of robustness, and relates that measure to Vidal and Tarrach's
measure.  The novel measure and its properties will be of especial
interest in applications to gate robustness.

%
%
Let $\rho$ be a quantum state of a bipartite system, $AB$, and let
$\sigma$ be a state of $AB$.  Vidal and Tarrach~\cite{Vidal99b} define
the \emph{robustness of $\rho$ relative to $\sigma$}, $R(\rho \|
\sigma)$, to be the smallest non-negative number $t$ such that the
state
\begin{eqnarray} \label{eq:relative_robustness}
  \frac{1}{1+t} \rho + \frac{t}{1+t} \sigma
\end{eqnarray}
is separable.  Equivalently, we can define $R(\rho \| \sigma)$ to be
the smallest non-negative number $t$ such that $\rho+t \sigma$ is
separable; this latter definition in terms of unnormalized quantum
states will frequently be useful in later work.  Note
that~\cite{Vidal99b} specify that $\sigma$ be separable; however, we
will find it convenient to extend the definition to non-separable
$\sigma$ also, specifying that $R(\rho \| \sigma) \equiv +\infty$ if no
value of $t$ exists such that the state in
Eq.~(\ref{eq:relative_robustness}) is separable.  At first sight one
is tempted to ask why we choose this definition for the robustness,
and not the related quantity
\begin{eqnarray} \label{eq:p-robust}
 \min \{p : p \geq 0, (1-p) \rho + p \sigma 
\mbox{ is separable} \}.
\end{eqnarray}
This latter definition has a more obvious physical interpretation as
the minimal probability with which $\sigma$ can be mixed with $\rho$
to obtain a separable state.  It follows from the definitions that the
quantity of Eq.~(\ref{eq:p-robust}) is equal to $R(\rho \|
\sigma)/(1+R(\rho \| \sigma))$.  The reason we do not work with the
quantity of Eq.~(\ref{eq:p-robust}), despite its apparently more
compelling physical interpretation, is that the robustness defined in
Eq.~(\ref{eq:relative_robustness}) has useful and easy-to-prove
convexity properties not satisfied by Eq.~(\ref{eq:p-robust}), namely,
$R(\rho \| \sigma)$ is convex in both the first and the second entry.

%
%
A special case of $R(\rho \| \sigma)$ of particular interest is the
\emph{random robustness}, defined to be the robustness of $\rho$
relative to the maximally mixed state $I/d_A d_B$.  We denote the
random robustness of a state $\rho$ by $R_r(\rho) \equiv R(\rho \|
I/d_A d_B)$.  Vidal and Tarrach~\cite{Vidal99b} found a useful formula
for the random robustness of a pure state $\psi$ of $AB$ in terms of a
Schmidt decomposition $\psi = \sum_j \psi_j |j\ra |j\ra$ with ordered
Schmidt coefficients $\psi_1 \geq \psi_2 \geq \ldots \geq 0$:
\begin{eqnarray} \label{eq:random-robustness}
R_r( \psi ) = \psi_1 \psi_2 d_A d_B.
\end{eqnarray}

%
%
So far we have discussed the robustness of a state $\rho$
\emph{relative} to another fixed state $\sigma$.  We now define
\emph{the robustness} of $\rho$, $R(\rho)$, to be the \emph{minimum}
relative robustness $R(\rho \| \sigma)$ over all separable $\sigma$.
Thus, the robustness of $\rho$ is a measure of how much local noise
can be mixed with $\rho$ before it becomes separable.

%
%
We have defined three notions of robustness for quantum states,
$R(\rho \| \sigma), R_r(\rho)$, and $R(\rho)$.  All three of these
definitions have assumed that $\rho$ is a state of a \emph{bipartite}
quantum system, $AB$.  However, robustness is easily extended to more
than two parties, and it is convenient to have a notation to express
the extended notion.  Suppose, for example, that $\rho$ and $\sigma$
are states of a tripartite system $ABC$.  Then $R^{A:B:C}(\rho \|
\sigma)$ is defined to be the minimal value of $t$ such that $\rho +
t\sigma$ is separable with respect to $A:B:C$.

%
%
Of course, a many-party quantum system can be decomposed in many
different ways, by grouping subsystems together.  So, for example, we
can define a notion of robustness, $R^{A:BC}(\rho\| \sigma)$, when
system $B$ and $C$ are grouped together.  More explicitly,
$R^{A:BC}(\rho\| \sigma)$ is defined to be the minimal value of $t$
such that $\rho+t\sigma$ is separable with respect to $A:BC$.

%
%
These examples may be extended in a natural way to the random
robustness and robustness, as well as to the case where more systems
are present, and to more complicated groupings of subsystems.  Most of
our work concerns two-party robustness, and so we usually do not
explicitly include superscripts in expressions like $R^{A:B}(\rho)$.

%
%
The robustness has many useful properties, which are explored in
detail in~\cite{Vidal99b}.  We mention just a few of the more striking
properties here.  The robustness is invariant under local unitary
operations.  Moreover, it is an entanglement monotone, that is, cannot
be increased under local operations and classical communication.  It
is also a convex function of $\rho$.  As for the random robustness,
Vidal and Tarrach~\cite{Vidal99b} have obtained an elegant formula for
the robustness in the special case of a pure state, $\psi$, of a
bipartite system, $AB$,
\begin{eqnarray}
  R(\psi) = \left( \sum_j \psi_j \right)^2-1,
\end{eqnarray}
where $\psi_j$ are the Schmidt coefficients for $\psi$.  In the course
of their proof, Vidal and Tarrach explicitly construct a state,
$\sigma_\psi$, such that $|\psi\rangle \langle\psi| + R(\psi) \sigma_\psi$ 
is separable.
$\sigma_\psi$ may be expressed in terms of the Schmidt decomposition
$\psi = \sum_j \psi_j |j\rangle |j\rangle$ by
\begin{eqnarray} \label{eq:VT-const}
  \sigma_\psi = \frac{1}{R(\psi)}
  \sum_{k \neq l} \psi_k \psi_l |k\rangle \langle k| \otimes |l\rangle
  \langle l|.
\end{eqnarray}

%
%
In the definition of robustness we mixed $\rho$ with a
\emph{separable} quantum state, $\sigma$, trying to determine what
minimal level of mixing will produce separability.  Another natural
definition of robustness would allow $\sigma$ to range over
\emph{arbitrary} density matrices, not just separable density
matrices.  That is, we can define $R_g(\rho) \equiv \min_\sigma R(\rho
\| \sigma)$, where the $g$ subscript indicates that we are minimizing
\emph{globally} over all possible density matrices $\sigma$.

%
%
How are $R_g(\rho)$ and $R(\rho)$ related?  It is clear from the
definitions that $R_g(\rho) \leq R(\rho)$.  We will prove that the
reverse inequality is also true when $\rho = \psi$ is a pure state, so
we have 
\begin{eqnarray} \label{eq:identity}
R_g(\psi) = R(\psi) = \left( \sum_j \psi_j \right)^2-1.
\end{eqnarray}
We do not know whether $R_g(\rho) = R(\rho)$ in general.  To complete
the proof of Eq.~(\ref{eq:identity}), we show that if there exists a
density operator $\sigma$ such that
\begin{eqnarray} \label{eq:sep-state}
 \psi + t \sigma
\end{eqnarray}
is separable, then $t \geq (\sum_j \psi_j)^2-1$.  (Our proof both
extends and simplifies a similar proof in~\cite{Vidal99b} for the
robustness $R(\rho)$.)

%
%
The proof is based on the positive partial transpose criterion of
Peres~\cite{Peres96b}.  Let us denote the partial transpose on systems
$A$ and $B$ by $T_A$ and $T_B$, respectively.  Then the positive
partial transpose criterion implies that if the state of
Eq.~(\ref{eq:sep-state}) is separable, then
\begin{eqnarray} \label{eq:sep-state-2}
  0 \leq \psi^{T_B} + t \, \sigma^{T_B},
\end{eqnarray}
where $\leq$ indicates an operator inequality, that is, we are saying
that the operator on the right-hand side is positive.

%
%
We now use Eq.~(\ref{eq:sep-state-2}) to deduce a lower bound on $t$.
To do this we introduce an operator, $M$, defined by $M \equiv
I-\swap$, where $\swap \equiv \sum_{jk} |j\rangle \langle k| \otimes
|k\rangle \langle j|$ is the linear operator interchanging states of
system $A$ and system $B$.  Note that $M$ is positive, since $\swap^2
= I$ implies that $\swap$ has eigenvalues $\pm 1$, and thus $M$ is a
diagonalizable operator with eigenvalues $0$ and $2$.

Since the trace of a product of two positive operators is
non-negative, it follows from Eq.~(\ref{eq:sep-state-2}) that
\begin{eqnarray} 
  0 \leq \tr\left( M\psi^{T_B} \right) + t \, \tr \left( M \sigma^{T_B}\right).
\end{eqnarray}
Using a little algebra and the observation that for any two operators,
$K$ and $L$, $\tr(KL^{T_B}) = \tr(K^{T_A}L)$, the previous equation
may be rewritten
\begin{eqnarray} \label{eq:ra-bound}
   -\tr\left( M^{T_A}\psi \right) \leq t \, \tr \left( M^{T_A} \sigma \right).
\end{eqnarray}
Direct calculation shows that
\begin{eqnarray} \label{eq:M-pt}
  M^{T_A} = I-|\alpha\rangle \langle \alpha|,
\end{eqnarray}
where $|\alpha\rangle \equiv \sum_j |j\rangle |j\rangle$ is the
(unnormalized) maximally entangled state.  Using Eq.~(\ref{eq:M-pt})
it follows that $\tr\left( M^{T_A}\psi \right) = 1-(\sum_j \psi_j)^2$
and $\tr \left( M^{T_A} \sigma \right) \leq \tr(\sigma) = 1$.
Substituting these results into Eq.~(\ref{eq:ra-bound}) gives
\begin{eqnarray}
  \left( \sum_j \psi_j \right)^2-1 \leq t,
\end{eqnarray}
which was the desired bound.

\section{Robustness of quantum gates}
\label{sec:results}

%
%
We now extend state robustness to \emph{quantum gates}.  Suppose
${\cal E}$ and ${\cal F}$ are trace-preserving quantum operations on
a composite system $AB$.  Then we define the \emph{robustness of
  ${\cal E}$ relative to ${\cal F}$}, $R({\cal E}\| {\cal F})$, to be
the minimum value of $t$ such that
\begin{eqnarray}
  \frac{1}{1+t} {\cal E}+\frac{t}{1+t} {\cal F}
\end{eqnarray}
is separable.  Equivalently, $R({\cal E}\| {\cal F})$ can be defined
to be the minimal value of $t$ such that ${\cal E}+t{\cal F}$ is
separable.  Applying the operation-separability theorem we immediately
find the useful formula
\begin{eqnarray} \label{eq:op-state-robustness}
  R({\cal E}\| {\cal F}) = R^{R_AA:BR_B}
  (\rho({\cal E}) \| \rho({\cal F})).
\end{eqnarray}

%
%
Just as for quantum states, the notion of gate robustness extends in a
natural way to systems of more than two parties, and we use notations
analogous to those introduced earlier, such as $R^{A:B:C}({\cal E}\|
{\cal F})$ and $R^{A:BC}({\cal E}\| {\cal F})$, to describe this
scenario.  Note that these notations will also be extended in a
natural way to the random robustness and robustness of a quantum gate,
as defined below.  As for quantum states, when identifying
superscripts are omitted we assume that the quantum gate in question
acts on a bipartite system, $AB$.

%
%
Motivated by several different classes of noise commonly occurring in
physical systems, we now use the notion of relative gate robustness to
define and study several different measures of robustness for quantum
gates.  First is the \emph{random robustness}, which we define and
study in Sec.~\ref{subsec:r-robustness}.  Also in this subsection, we
use results on the random robustness to place bounds on the threshold
for quantum computation.  Two other measures of robustness are the
\emph{separable robustness} and the \emph{global robustness}, which we
define in Sec.~\ref{subsec:other-robustness}, and use to prove bounds
on the threshold for quantum computation.  Our results on these
measures of robustness are less complete, and so our discussion is
more limited.

\subsection{Random robustness of quantum gates}
\label{subsec:r-robustness}

\subsubsection{Definition and basic properties}

%
%
The \emph{random robustness of ${\cal E}$}, $R_r({\cal E})$, is
defined to be equal to the robustness of ${\cal E}$ relative to the
completely depolarizing channel, ${\cal D}(\rho) = I/d_A d_B$ for all
states $\rho$ of system $AB$:
\begin{eqnarray}
R_r({\cal E}) \equiv R({\cal E} \| {\cal D}).
\end{eqnarray}
The random robustness is especially interesting because it measures
the robustness of ${\cal E}$ against complete randomization of systems
$A$ and $B$.  Another way of stating this is to imagine that we are
applying the operation ${\cal E}$ with probability $1-p$, and
randomizing the systems $A$ and $B$ with probability $p$.  Then the
threshold probability at which this gate crosses the
separable-inseparable threshold is:
\begin{eqnarray}
  \frac{R_r({\cal E})}{1+R_r({\cal E})}.
\end{eqnarray}
From Eq.~(\ref{eq:op-state-robustness}) we see that the random
robustness for an operation is related to the random robustness of a
state by
\begin{eqnarray} \label{eq:op-random-robustness}
  R_r({\cal E}) = R_r^{R_AA:BR_B}(\rho({\cal E})).
\end{eqnarray}
Specializing to the case where ${\cal E}$ is a unitary quantum
operation, $U$, we see that $R_r(U) = R_r^{R_AA:BR_B}(\rho(U))$.  However,
$\rho(U)$ is a pure state.  We showed earlier that $\rho(U)$ has
Schmidt coefficients $u_j/\sqrt{d_A d_B}$, where $u_j$ are the Schmidt
coefficients of $U$.  This observation, together with
Eqs.~(\ref{eq:op-random-robustness}) and~(\ref{eq:random-robustness})
implies the formula
\begin{eqnarray} \label{eq:unitary-random-robustness}
R_r(U) = d_A d_B u_1 u_2,
\end{eqnarray}
where we order the Schmidt coefficients of $U$ so that $u_1 \geq u_2
\geq \ldots \geq 0$.  (Note that in deriving this equation, we have
replaced $d_A$ by $d_A^2$, and $d_B$ by $d_B^2$ in
Eq.~(\ref{eq:random-robustness}), since we are working with robustness
for the $R_A A:BR_B$ system.)

%
%
It is, perhaps, not immediately clear what the physical relevance of
the random robustness is.  After all, in real physical systems, the
effects of noise on a quantum gate will not usually be to simply mix
in some depolarization, together with the gate.  Despite this, there
is still a very good physical reason to be interested in the random
robustness.  The reason is that, as we show in more explicit detail
below, the random robustness can be used to analyze the particular
noise models which have been used in estimating bounds on the threshold
for quantum computation.  In turn, it has been
argued~\cite{Aharonov99a,Gottesman97a,Kitaev97a,Knill98a,Preskill98b}
that by analyzing and correcting for the effects of noise in those
\emph{particular} models, it is possible to make general statements
about a wide class of physically reasonable noise models.  Thus,
although the physical scenario considered in the definition of the
random robustness appears rather specialized, it will enable insight
into much more general physical situations.

%
%
As an example, we may ask how robust the controlled-{\sc not} is
against the effects of depolarizing noise?  The controlled-{\sc not}
has Schmidt decomposition~\cite{Nielsen02a} $\sqrt{2} |0\rangle
\langle 0| \otimes I/\sqrt{2} + \sqrt{2} |1\rangle \langle 1| \otimes
X/\sqrt{2}$, so Eq.~(\ref{eq:unitary-random-robustness}) implies that
$R_r(\mbox{\sc cnot}) = 8$.  Interestingly, we can also show that the
controlled-{\sc not} is the most robust two-qubit gate.  To see this,
note that unitarity of $U$ implies that the Schmidt coefficients $u_j$
satisfy $\sum_j u_j^2 = d_A d_B$, and thus $u_1^2+u_2^2 \leq d_A d_B$.
It follows from this observation and
Eq.~(\ref{eq:unitary-random-robustness}) that $R_r(U) \leq d_A^2
d_B^2/2$, and thus no two-qubit unitary gate can have random
robustness greater than $8$, which is the random robustness of the
controlled-{\sc not}.  These results are worth highlighting as a
proposition:

\begin{proposition}
For any quantum operation ${\cal E}$, $R_r({\cal E})\leq
d_A^2d_B^2/2$.  If $d_A=d_B=2$ then $R_r({\cal E})\leq R_r({\sc
CNOT})=8$.
\end{proposition}

%
%
The random robustness has many physically interesting properties.
Below we list six easily-proved properties, before discussing in more
depth two less easily-proved properties.  Our discussion of these
properties is, in part, motivated by the framework of ``dynamic
strength'' measures introduced in~\cite{Nielsen02a}, although the
properties we discuss are interesting independent of that motivation.
In~\cite{Nielsen02a} it was argued that these properties, especially
the property of \emph{chaining}, discussed below, are essential if a
measure can be said to quantify the strength of a quantum dynamical
operation as a physical resource.  By showing that these properties
are satisfied, we thus show that the random robustness is a good
measure of dynamic strength.

\begin{enumerate}

\item \textbf{Non-negativity and locality:} $R_r({\cal E}) \geq 0$
  with equality if and only if ${\cal E}$ is a separable quantum
  operation.
  
\item \textbf{Local unitary invariance:} If ${\cal U}_A,{\cal U}_B,
  {\cal V}_A,{\cal V}_B$ are all local unitary quantum operations,
  with the system being acted on indicated by the subscript, then
  \begin{eqnarray}
    R_r(({\cal U}_A \otimes {\cal U}_B) \circ {\cal E} 
    \circ ({\cal V}_A \otimes {\cal V}_B)) = R_r({\cal E}).
  \end{eqnarray}
  
\item \textbf{Exchange symmetry:} $R_r({\cal E}) = R_r(\swap \circ
  {\cal E} \circ \swap)$, that is, the random robustness is not
  affected if we interchange the role of the systems.
  
\item \textbf{Time-reversal invariance:} For a unitary, $U$, $R_r(U) =
  R_r(U^\dagger)$.
  
\item \textbf{Convexity:} The random robustness $R_r({\cal E})$ is
  \emph{convex} in ${\cal E}$. 
  
%
  
\item \textbf{Reduction:} Suppose a trace-preserving quantum operation
  ${\cal E}$ acting on $AB$ is obtained from a trace-preserving quantum
  operation ${\cal F}$ acting on $ABC$ as follows:
  \begin{eqnarray}
    {\cal E}(\rho_{AB}) = \tr_C\left[ {\cal F}(\rho_{AB} \otimes \sigma_C)
      \right],
  \end{eqnarray}
  for some fixed state $\sigma_C$ of system $C$.  Then the random
  robustness satisfies the reduction property, namely,
  $R_r^{A:B}({\cal E}) \leq R_r^{A:BC}({\cal F})$.

\end{enumerate}

The random robustness satisfies two other physically interesting
properties that are more difficult to prove.  First of all, the random
robustness is \emph{continuous} in ${\cal E}$.  Physically, this is
self-evident: making a small change in ${\cal E}$ should not too
drastically affect its robustness against the effects of noise.  We
now prove a quantitative form of this statement for unitary gates.

\begin{proposition}[Continuity of random robustness]
  Let $U$ and $V$ be unitary gates acting on a system $A$ of dimension
  $d_A$, and a system $B$ of dimension $d_B$.  Then
\begin{eqnarray}
  |R_r(U)-R_r(V)| \leq d_M d_A^3 d_B^3 \| U-V\|^2,
\end{eqnarray}
where $d_M \equiv \min(d_A,d_B)$.
\end{proposition}

\textbf{Proof:} Let $u_j$ and $v_j$ be the ordered Schmidt
coefficients of $U$ and $V$, respectively.  From
Eq.~(\ref{eq:unitary-random-robustness}),
\begin{eqnarray}
|R_r(U)-R_r(V)| & = & d_Ad_B |u_1u_2-v_1v_2| \nonumber \\
& = & d_A d_B |(u_1-v_1)u_2+v_1(u_2-v_2) | \nonumber \\
& \leq & d_A d_B |u_1-v_1| |u_2| + |v_1| |u_2-v_2| \nonumber \\
& \leq & d_A^2 d_B^2 \left( |u_1-v_1| + |u_2-v_2| \right) \nonumber \\
& \leq & d_A^2 d_B^2 \sum_j |u_j-v_j|. \label{eq:r-r-inter}
\end{eqnarray}

The second part of the proof is to observe that by the Cauchy-Schwartz
inequality,
\begin{eqnarray}
  \sum_j |u_j-v_j| & \leq & d_M \sum_j (u_j^2+v_j^2-2u_jv_j) \\
  & = & 2d_M d_A d_B \left(1-\frac{\sum_j u_j v_j}{d_A d_B} \right).
\end{eqnarray}
Applying Proposition~\ref{prop:Schmidt-continuity} we obtain
\begin{eqnarray}
  \sum_j |u_j-v_j| \leq d_M d_A d_B \| U - V \|^2
\end{eqnarray}
Combining with Eq.~(\ref{eq:r-r-inter}) gives the result.  
\qed

%
%
Another physically interesting question is to ask how the random
robustness of a gate ${\cal E}_1 \circ {\cal E}_2$ composed of quantum
gates ${\cal E}_1$ and ${\cal E}_2$ relates to the random robustness
of the individual gates.  The following proposition bounds the
random robustness of the combined operation:
\begin{proposition}[Chaining for random robustness]{\label{prop:chaining}}
  Let ${\cal E}_1$ be a doubly stochastic quantum operation, that is,
  a quantum operation which is both trace-preserving and unital (i.e.
  ${\cal E}_1(I) = I$), and let ${\cal E}_2$ be an arbitrary
  trace-preserving quantum operation.  Then
  \begin{eqnarray}
    R_r({\cal E}_1 \circ {\cal E}_2) \leq R_r({\cal E}_1)+R_r({\cal E}_2)
    +R_r({\cal E}_1)R_r({\cal E}_2).
  \end{eqnarray}
\end{proposition}

%
%
Note that unitary operations are trace-preserving and unital, so the
proposition is true when ${\cal E}_1$ and ${\cal E}_2$ are unitary.
There is an equivalent way of phrasing Proposition~\ref{prop:chaining}
that is physically more intuitive.  Suppose we define
\begin{eqnarray}
 C_r({\cal E}) \equiv \ln(1+R_r({\cal E})).
\end{eqnarray}
Then $C_r({\cal E})$ is monotonically related to the random robustness
of ${\cal E}$, and thus can be thought of as carrying the same
qualitative information about the robustness of the gate.  Simple
algebra shows that the conclusion of Proposition~\ref{prop:chaining}
may be recast in the form
\begin{eqnarray}
  C_r({\cal E}_1 \circ {\cal E}_2) \leq C_r({\cal E}_1)+C_r({\cal E}_2).
\end{eqnarray}
The simplicity and clarity of this form may, perhaps, make it more
useful in some circumstances.

\textbf{Proof:}
By definition of the random robustness, the quantum operations
\begin{eqnarray}
  & & {\cal E}_1+R_r({\cal E}_1){\cal D}, \hspace{3mm} \mbox{ and} 
\label{eq:chaining-inter-1} \\
 & & {\cal E}_2+R_r({\cal E}_2){\cal D}
\label{eq:chaining-inter-2}
\end{eqnarray}
are separable quantum operations.  Furthermore, since the composition
of two separable quantum operations is separable, and ${\cal E}_1
\circ {\cal D} = {\cal D} \circ {\cal E}_2 = {\cal D} \circ {\cal D} =
{\cal D}$ (using the unitality of ${\cal E}_1$), we can compose the
operations of Eqs.~(\ref{eq:chaining-inter-1})
and~(\ref{eq:chaining-inter-2}) to see that
\begin{eqnarray}
 & & {\cal E}_1 \circ {\cal E}_2 +
\left( R_r({\cal E}_1)+R_r({\cal E}_2)
    +R_r({\cal E}_1)R_r({\cal E}_2)\right){\cal D} \nonumber \\
& &
\end{eqnarray}
is separable, and thus
\begin{eqnarray}
  R_r({\cal E}_1 \circ {\cal E}_2) \leq R_r({\cal E}_1)+R_r({\cal E}_2)
  + R_r({\cal E}_1)R_r({\cal E}_2),
\end{eqnarray}
as required. \qed

\subsubsection{Random robustness and the threshold for quantum
  computation}\label{sec:rr-threshold}

%
%
Suppose we are trying to do fault-tolerant quantum computation using
single-qubit gates and some entangling two-qubit unitary gate, $U$.
$U$ might be the controlled-{\sc not} gate; it can also be any other
entangling two-qubit gate, at least in
principle~\cite{Brylinski02a,Bremner02a}, and still be capable of
universal quantum computation when assisted by single-qubit gates.
Suppose furthermore that the $U$ gates are afflicted with noise of a
special type, namely, immediately after a gate acts, each qubit is
independently depolarized with probability $p$.  Let ${\cal U}(\rho)
\equiv U\rho U^\dagger$ denote the quantum operation corresponding to
$U$.  Then the quantum operation describing this noise process is
\begin{eqnarray}
  {\cal E}(\rho) & = & (1-p)^2 {\cal U}(\rho) + p(1-p) 
  ({\cal D}\otimes {\cal I})\circ{\cal U} (\rho) \nonumber \\
  & & + p(1-p)({\cal I} \otimes {\cal D})\circ{\cal U}  (\rho)+p^2 
  ({\cal D} \otimes {\cal D})\circ{\cal U} (\rho). \nonumber \\
  & & \label{eq:noise-process}
\end{eqnarray}
Note that $({\cal D} \otimes {\cal D})\circ{\cal U}(\rho) = ({\cal
  D} \otimes {\cal D})(\rho)$, so this expression can be simplified to
\begin{eqnarray}
  {\cal E}(\rho) & = & (1-p)^2 {\cal U}(\rho) + p(1-p) 
  ({\cal D}\otimes {\cal I})\circ{\cal U} (\rho) \nonumber \\
  & & + p(1-p)({\cal I} \otimes {\cal D})\circ{\cal U}  (\rho)+p^2 
  ({\cal D} \otimes {\cal D}) (\rho). \nonumber \\
  & & \label{eq:threshold-inter}
\end{eqnarray}
This expression cannot immediately be analyzed using our expressions
for the random robustness of a gate, due to the two terms in which a
single qubit is depolarized.  Fortunately, we can simplify the
analysis by showing that these terms are always
separability-preserving, that is, $({\cal D}\otimes {\cal I}) \circ
{\cal U} $ and $({\cal I} \otimes {\cal D}) \circ {\cal U}$ are both
in {\bf SP}.  This holds because for any $\rho$, $({\cal D}\otimes
{\cal I})\circ{\cal U}(\rho)=\frac{I}{d_A}\otimes \tr_A U\rho U^\dag$, which
is manifestly separable, and a similar result holds for $({\cal I} \otimes
{\cal D})\circ{\cal U}$.  Note that such gates may not be 
separable: for example $({\cal D}\otimes {\cal I}) \circ \swap +
({\cal I} \otimes {\cal D}) \circ \swap$ is separability-preserving,
but not separable.

From this observation, and Eq.~(\ref{eq:threshold-inter}), it follows
that ${\cal E}$ is in {\bf SP} if $(1-p)^2 {\cal U} + p^2({\cal D}
\otimes {\cal D})$ is separable.  Comparing with the earlier results
on random robustness, we see that this becomes true when $p^2/(1-p)^2
= R_r(U)=8$.  We see that ${\cal E}$ will be
separability-preserving when:
\begin{eqnarray}
  p \geq \frac{R_r(U)-\sqrt{R_r(U)}}{R_r(U)-1} = \frac{8-\sqrt{8}}{7}
  \approx 0.74,
\end{eqnarray}
and thus, when this condition is satisfied, the quantum computation
may be efficiently simulated on a classical computer.  If we assume,
as is usually done, that quantum computers may not be efficiently
simulated on a classical computer, then it follows that the threshold
for quantum computation is guaranteed to be less than $0.74$.

%
%
In their work on obtaining upper bounds for the threshold, Aharonov
and Ben-Or~\cite{Aharonov96b} considered a similar model of quantum
computation, in which each qubit is independently dephased after each
quantum gate. The main difference between their model and ours is that
we have used depolarizing, rather than dephasing noise.  Which of
these more accurately describes the noise occurring in a real physical
system depends, of course, upon the physical system in question.
Aharonov and Ben-Or's obtained an upper bound of $p_{\rm th} < 0.97$;
of course, this cannot be directly compared to our upper bound, since
the noise models are different.

%

\subsection{Robustness against more general noise}
\label{subsec:other-robustness}

\subsubsection{Definitions and general results}
%
%
Depolarization is only one of the many kinds of noise that can afflict
a quantum gate.  Other classes of noise motivate other measures of
robustness for quantum gates.  We now introduce two more measures of
robustness, based on two particularly natural classes of noise.  The
first measure is the \emph{separable robustness}, which measures the
resilience of the gate against separable noise.  The separable
robustness, $R_s({\cal E})$, is defined to be the minimum relative
robustness $R({\cal E} \| {\cal F})$ over all separable,
trace-preserving quantum operations ${\cal F}$.  The second measure is
the \emph{global robustness}, which measures the resilience of the
gate against arbitrary noise.  The global robustness, $R_g({\cal E})$,
is defined to be the minimum relative robustness $R({\cal E} \| {\cal
  F})$ over all trace-preserving quantum operations ${\cal F}$.

%
%
\emph{A priori}, it is apparent that $R_g({\cal E}) \leq R_s({\cal
  E})$, but it is not clear whether or not the two quantities are
equal.  Furthermore, the gate robustnesses may be related to state
robustness by the following inequalities:
\begin{eqnarray} 
  R(\rho({\cal E})) \leq R_s({\cal E}), \label{eq:ineq1} \\
  R(\rho(U)) \leq R_g(U). \label{eq:ineq2}
\end{eqnarray}
To see the first of these inequalities, note that ${\cal E}+R_s({\cal
  E}) {\cal F}$ is separable, for some separable quantum operation
${\cal F}$.  It follows that
\begin{eqnarray}
  \rho({\cal E}+R_s({\cal E}) {\cal F}) = \rho({\cal E})+R_s({\cal E})
  \rho({\cal F})
\end{eqnarray}
is a separable quantum state.  Since $\rho({\cal F})$ is separable,
Eq.~(\ref{eq:ineq1}) follows from the definition of $R(\rho({\cal E}))$.
The proof of Eq.~(\ref{eq:ineq2}) follows similar lines, but also
makes use of the fact, noted in Eq.~(\ref{eq:identity}), that
$R_g(\psi) = R(\psi)$ for any pure state $\psi$.

%
%
Do the inequalities~(\ref{eq:ineq1}) and~(\ref{eq:ineq2}) hold with
equality?  We do not know the answer to this question, but suspect
that the answer is, in general, ``no'', in both cases.  Our reasoning
for this suspicion is as follows. Recall from
Sec.~\ref{sec:separable}, in particular, Theorem~\ref{thm:sep-char},
that not all separable states can be written as $\rho({\cal F})$ for
some separable quantum operation ${\cal F}$.  Recall also the
construction, Eq.~(\ref{eq:VT-const}), used in finding the separable
$\sigma_\psi$ which minimizes $R(\psi \| \sigma_\psi)$.  Using this
construction, it is not difficult to find examples of unitary $U$ for
which the separable state $\sigma_{\rho(U)}$ does not correspond to
any trace-preserving, separable quantum operation, as characterized in
Theorem~\ref{thm:sep-char}.

%
%
Fortunately, there is a large and interesting class of gates for which
the inequalities~(\ref{eq:ineq1}) and~(\ref{eq:ineq2}) hold with
equality.  This class includes the controlled-{\sc not} and {\sc swap}
gates.

\begin{theorem}\label{thm:unital-gate}
  Let $U$ be a bipartite unitary gate acting on systems $A$ and $B$
  with dimensions $d_A$ and $d_B$.  Assume $U$ has Schmidt decomposition
  \begin{eqnarray}
    U = \sum_j u_j A_j \otimes B_j,
  \end{eqnarray}
  where the $A_j$ satisfy $A_j A_j^\dagger = I/d_A$ and the
  $B_j$ satisfy $B_j B_j^\dagger = I/d_B$.  That is, the
  $A_j$ and $B_j$ are all proportional to unitary operators.  Then
  \begin{eqnarray} \label{eq:general-gate-robustness}
    R_g(U) = R_s(U) = R(\rho(U)) = \frac{\left( \sum_j u_j\right)^2}{d_Ad_B}-1.
  \end{eqnarray}
  Furthermore, the quantum operation ${\cal F}$ defined by
  \begin{eqnarray} \label{eq:worst-noise}
    {\cal F}(\rho) \equiv \frac{\sum_{k\neq l} u_k u_l \left( A_k\otimes
        B_l\right) \rho \left( A_k^\dagger \otimes B_l^\dagger\right)}{
      \sum_{k\neq l} u_k u_l }
  \end{eqnarray}
  is an instance of the type of noise against which $U$ is least
  robust.  That is, ${\cal F}$ is trace-preserving, and ${\cal
    U}+R(\rho(U)) {\cal F}$ is separable.  Finally, ${\cal F}$ is
  manifestly separable.  Indeed, ${\cal F}$ can be implemented by
  local operations and classical communication.
\end{theorem}

%
%
The application of the theorem of most interest for us is the {\sc
  cnot}.  It is not necessarily obvious that the {\sc cnot} has a
Schmidt decomposition with the properties required by the theorem;
after all, we earlier wrote the Schmidt decomposition for the {\sc
  cnot} as $\sqrt{2} |0\rangle \langle 0| \otimes I/\sqrt 2+ \sqrt 2
|1\rangle \langle 1| \otimes X / \sqrt 2$, and this is not of the
required form.  However, while the Schmidt co-efficients are unique,
the operators appearing in the Schmidt decomposition may not be
unique, when two or more of the co-efficients are degenerate.  It
turns out that there is an alternative form of the Schmidt
decomposition for the {\sc cnot} which is of the right form.  This
follows, for example, from Proposition~4 of~\cite{Nielsen02a}, and can
also be verified directly, with a little algebra.  The explicit form
is not particularly illuminating, so we omit it here.

Eq.~(\ref{eq:general-gate-robustness}) now tells us that
$R_g(\mbox{\sc cnot}) = R_s(\mbox{\sc cnot}) = 1$.  Comparing with the
random robustness, $R_r(\mbox{\sc cnot}) = 8$, we see that the {\sc
  cnot} is substantially less robust against general noise than
depolarizing noise.  From Eq.~(\ref{eq:worst-noise}) we deduce that the
quantum operation
\begin{eqnarray}
  {\cal F}(\rho) & = & \left( |0\rangle \langle 0| \otimes I \right)
  \rho \left( |0\rangle \langle 0| \otimes I \right) \nonumber \\
  & & + \left( |1\rangle \langle 1| \otimes X \right) \rho
  \left( |1\rangle \langle 1| \otimes X \right)
\end{eqnarray}
is an example of a noise process such that $\mbox{\sc cnot}+{\cal F}$
is a separable quantum operation.  Note, furthermore, that ${\cal F}$
can be implemented via local operations and classical communication,
by measuring the first qubit, and then conditionally applying $I$ or
$X$ to the second qubit, depending upon whether the outcome of the
measurement was zero or one.  It is interesting that ${\cal F}$ thus
corresponds to a classical {\sc cnot} operation.

\textbf{Proof:} We already know that $R(\rho(U)) \leq R_g(U) \leq
R_s(U)$, so it suffices to prove that $R_s(U) \leq R(\rho(U))$. To
prove this, we use the construction of Vidal and Tarrach,
Eq.~(\ref{eq:VT-const}), to see that $\rho(U) +
R(\rho(U))\sigma_{\rho(U)}$ is separable, where
\begin{eqnarray}
  & & \sigma(\rho(U)) = \frac{1}{R(\rho(U))} 
  \sum_{k\neq l} u_k u_k |k\rangle \langle k| \otimes |l\rangle \langle l|;
\\
& & |k\rangle \equiv (I_{R_A} \otimes A_k) |\alpha\rangle; \,\,\,\,
  |l\rangle \equiv (B_l\otimes I_{R_B}) |\beta\rangle.
\end{eqnarray}
Using the fact that the $A_k$ and $B_l$ are proportional to unitary
operations, a calculation shows that $\tr_{AB}(\sigma_{\rho(U)})$ is a
completely mixed, separable state.  By Theorem~\ref{thm:sep-char} we
conclude that there exists a trace-preserving, separable quantum
operation ${\cal F}$ such that $\rho({\cal F}) = \sigma_{\rho(U)}$.
(Another way of seeing this is to directly verify that ${\cal F}$ as
defined by Eq.~(\ref{eq:worst-noise}) satisfies $\rho({\cal F}) =
\sigma_{\rho(U)}$).  Thus
\begin{eqnarray}
  \rho(U) + R(\rho(U))\rho({\cal F}) =   \rho\left[{\cal U} + R(\rho(U)) 
    {\cal F}\right]
\end{eqnarray}
is separable, whence ${\cal U} + R(\rho(U)){\cal F}$ is separable.  It
follows from the definition that $R_s(U) \leq R(\rho(U))$, which
completes the proof.
\qed

%
%
It is not difficult to verify that $R_s({\cal E})$ and $R_g({\cal E})$
satisfy properties similar to those satisfied by the random
robustness, and thus can be regarded as measures of dynamic strength.
The major difference is continuity: the lack of an explicit formula
for the separable and global robustness has prevented us from
obtaining quantitative continuity statements like those we
obtained for the random robustness, although it is still not difficult
to argue that both quantities are continuous.

\subsubsection{General robustness and the threshold for quantum
computation}

%
%
As with the random robustness, we can use $R_s$ and $R_g$ to obtain
bounds on the threshold for quantum computation.  The method for
obtaining a bound is similar.  Suppose we have a quantum computer
capable of arbitrary single-qubit gates and a single two-qubit gate,
$U$.  Then there exists ${\cal E}$ such that $U + R_g(U){\cal E}$ is
separable.  Now we choose the following noise model: whenever we apply
$U$, there is probability $p$ that instead ${\cal E}$ occurs.  If $p
\geq R_g(U)/(1+R_g(U))$ then this set of operations can be efficiently
simulated classically, and we therefore conclude that $p_{\text th}
\leq R_g(U)/(1+R_g(U))$.  Similar remarks apply for $R_s(U)$, only the
noise in that case is restricted to be separable.

%
%
Note that both these noise models are more adversarial, or
pessimistic, than the noise model in Sec.~\ref{sec:rr-threshold}, and
the threshold bounds are thus tighter.  In particular, these models
allow correlated two-qubit noise, while the earlier model assumes
independent noise on the two qubits.  Which model is more realistic
obviously depends upon which system a gate is implemented in.
However, we do expect correlated errors similar to those in the
present models to play a role in many real-world two-qubit gates, due
to interactions occurring during the gate.

%
%
The bounds obtained using~$R_s$ and~$R_g$ are, in general, tighter
than those obtained by studying $R_r$, as in
Sec.~\ref{sec:rr-threshold}.  However, without specific formulas for
$R_s(U)$ and $R_g(U)$ it is difficult to derive bounds on the
threshold without resorting to numerical calculation.  Fortunately,
if the only entangling gate available is of the form described by
Theorem~\ref{thm:unital-gate}, then we can calculate the optimal noise
process, and the corresponding robustness $R_s(U)=R_g(U)=R(\rho(U))$.
For example, for the {\sc cnot}, this gives the bound $p_{\text th}
\leq 1/2$ on the threshold, since $R_g(\text{\sc cnot}) =
R_s(\text{\sc cnot}) = 1$.

%
%
An alternative approach to proving bounds on the threshold is provided
by the following general bound on the robustness.  The bound says,
roughly, that if all two-qubit unitary gates are available, then
without loss of generality the worst noise is depolarizing noise.

\begin{theorem}
  For any trace-preserving quantum operation ${\cal E}$, $\max_U R(U
  \| {\cal E}) \geq \max_U R_r(U ) = d_A^2d_B^2/2$.  
\end{theorem}

As a corollary of the theorem, we see that if all one- and two-qubit
gates are available, but we don't make any assumptions about the
noise, the worst possible noise will be depolarizing noise, ${\cal D}
\otimes {\cal D}$, and the corresponding bound on the threshould is
$p_{\text th} \leq 8/9$.

\textbf{Proof:} Completely depolarizing noise can be represented as
applying a random unitary operation $V_k$ with probability $p_k$,
where each $V_k=V_k^A\otimes V_k^B$ is a product of local gates and
$\sum p_k V_k\rho V_k^\dag \propto I$ for any density operator $\rho$.
Thus ${\cal D}=\sum p_k {\cal V}_k$ where ${\cal V}_k(\rho)=V_k\rho
V_k^\dag$.  

Since $R$ is convex in the
second argument and ${\cal D}\circ{\cal E} = {\cal D}$ for any
operation ${\cal E}$, it follows that for any unitary $U$,
\begin{eqnarray}
R(U\|{\cal D}) & = & R(U\|{\cal D} \circ {\cal E})
=R\left( U \left\| \sum_k p_k {\cal V}_k \circ {\cal E}\right.\right)
\nonumber \\ & \leq &
\sum_k p_k R\left(U \| {\cal V}_k \circ {\cal E}\right)
= \sum_k p_k R(V_k^\dag U \| {\cal E}),  \nonumber \\
& & \label{eq:optimal-noise}
\end{eqnarray}
where the last equality follows from the fact that $V_k$ is a product
of local gates.

Let $R_0=\max_U R(U \| {\cal E}) = \max_U R(V_k^\dag U \| {\cal E})$.
Then Eq.~(\ref{eq:optimal-noise}) implies that $R(U\|{\cal D})\leq
\sum_k p_k R_0 = R_0$ for any $U$, so $\max_U R(U\|{\cal D})\leq
\max_U R(U\|{\cal E})$ for any trace-preserving operation ${\cal E}$.  \qed

%
%
We conclude with a result tying our techniques more closely to the
physical situation.  Suppose we are attempting to perform quantum
computation in the laboratory using a noisy gate, ${\cal E}$, meant to
approximate an ideal, unitary quantum gate, $U$.  $U$ is known
exactly, for it is a theoretical construct, and ${\cal E}$ has been
experimentally determined using quantum process
tomography~\cite{Chuang97a,Poyatos97a}.  For what values of $p$ is it
possible to find a trace-preserving quantum operation, ${\cal G}$,
such that ${\cal E} = p {\cal U}+(1-p){\cal G}$?  The answer to a
generalization of this question is provided by the following theorem:

\begin{theorem}
  Let ${\cal E}$ and ${\cal F}$ be trace-preserving quantum
  operations, and let $0 \leq p \leq 1$.  Then there exists a
  trace-preserving quantum operation ${\cal G}$ such that ${\cal E} =
  p{\cal F}+(1-p){\cal G}$ if and only if the support of $\rho({\cal
    F})$ is contained within the support of $\rho({\cal E})$, and
  \begin{eqnarray}
    p \leq \frac{1}{\lambda_1(\rho({\cal E})^{-1}\rho({\cal F}))},
  \end{eqnarray}
  where $\lambda_1(\cdot)$ denotes the largest eigenvalue, and the
  inverse is a generalized inverse if $\rho({\cal E})$ is not
  invertible.
\end{theorem}

The theorem is a straightforward consequence of the following theorem,
and the Jamiolkowski~\cite{Jamiolkowski72a} isomorphism between states
and operations.

\begin{theorem} \label{thm:MH}
  Let $\rho$ and $\sigma$ be density matrices, and let $0 \leq p \leq
  1$.  Then there exists a density matrix $\tau$ such that $\rho =
  p\sigma+(1-p)\tau$ if and only if the support of $\sigma$ is
  contained within the support of $\rho$, and
  \begin{eqnarray}
    p \leq \frac{1}{\lambda_1(\rho^{-1}\sigma)},
  \end{eqnarray}
  where $\lambda_1(\cdot)$ denotes the largest eigenvalue, and the
  inverse is a generalized inverse if $\rho$ is not
  invertible.
\end{theorem}

\textbf{Proof:} Suppose $\rho = p\sigma+(1-p)\tau$.  Since $\sigma$
and $\tau$ are positive, it is clear that the support of both $\sigma$
and $\tau$ must be contained within the support of $\rho$.  It will be
convenient to work in the vector space corresponding to the support of
$\rho$, so $\rho$ is invertible.  Since $\tau$ is positive, we have
$\rho \geq p\sigma$, as an operator inequality.  Pre- and
post-multiplying by $\rho^{-1/2}$ gives $I \geq p \rho^{-1/2} \sigma
\rho^{-1/2}$.  Comparing the largest eigenvalues of these two
operators gives the desired inequality.  The converse is proved by
running the argument backward.  \qed

\section{Conclusion}
\label{sec:conclusion}

%
%
We have defined several measures of the robustness of quantum gates
against the effects of noise, and used these measures to prove that
certain noisy quantum gate sets can be efficiently simulated on a
classical computer, even if the methods of fault-tolerant computation
are used.  Our results imply an upper bound on the threshold for
quantum computation, $p_{\rm th} \leq 0.5$.  A key component in
proving these results was a proof that any quantum computation
involving only separable quantum gates can be efficiently simulated on
a classical computer.  Furthermore, we have studied gate robustness as
a measure of the strength of a quantum operation, considered as a
physical resource, and shown that robustness satisfies many properties
such a strength measure is expected to have.

\acknowledgments 

Thanks to the quantum information theory group at the University of
Queensland for valuable discussions on gate robustness and
entanglement.  MAN thanks Michael Hall for enlightening and enjoyable
discussions which led to the proof of Theorem~\ref{thm:MH}.  AWH
thanks the Centre for Quantum Computer Technology at the University of
Queensland for its hospitality and acknowledges support from the NSA
and ARDA under Army Research Office contract number DAAD19-01-1-06.


\end{document}